# Adapting the Predator-Prey Game Theoretic Environment to Army Tactical Edge Scenarios with Computational Multiagent Systems

Derrik E. Asher[1], Erin Zaroukian[1], Sean L. Barton[1]

1. US Army Research Laboratory, Computational & Information Sciences Directorate

## Abstract

The historical origins of the game theoretic predator-prey pursuit problem can be traced back to Benda, et al., 1985 [1]. Their work adapted the predator-prey ecology problem into a pursuit environment which focused on the dynamics of cooperative behavior between predator agents. Modifications to the predator-prey ecology problem [2] have been implemented to understand how variations to predator [3] and prey [3-5] attributes, including communication [6], can modify dynamic interactions between entities that emerge within that environment [7-9]. Furthermore, the predator-prey pursuit environment has become a testbed for simulation experiments with computational multiagent systems [10-12]. This article extends the theoretical contributions of previous work by providing 1) additional variations to predator and prey attributes for simulated multiagent systems in the pursuit problem, and 2) military-relevant predator-prey environments simulating highly dynamic, tactical edge scenarios that Soldiers might encounter on future battlefields. Through this exploration of simulated tactical edge scenarios with computational multiagent systems, Soldiers will have a greater chance to achieve overmatch on the battlefields of tomorrow.

## Background

The predator-prey paradigm originally emerged from the field of ecology and was analyzed through a series of differential equations describing population dynamics among at least two species (often predator and prey) [13]. Over time, this evolved into a framework for investigating environmental and spatial contributions towards behavioral dynamics. From 1925 to 1966 modifications to the predator-prey model resulted in the emergence of functional predator behavior dependent on prey death rate and prey population density [14-16]. These modifications introduced the first inferred spatial component to the predator-prey paradigm. In 1985, the predator-prey model was extended to the problem of individual pursuit (i.e., focusing on individual interactions rather than population dynamics). This shift was aimed at exploring cooperative behavioral dynamics in multi-agent systems [1]. Since then, the predator-prey pursuit problem has begun to see use as a benchmark for testing multiagent algorithms, due to the inherent competitive and cooperative elements intrinsic to its design [17].

The typical predator-prey pursuit environment consists of multiple predators and a single prey moving around in a 2-D confined arena either discretely (one space at a time in either up, down, left, or right directions) or continuously (smooth continuous movement in any direction) for a fixed duration [7, 11, 12, 17-22]. The goals of the predators are in direct competition with those of the prey. The predators' shared goal is to come in contact with (continuous movement) or settle adjacent to (discrete movement) the prey, while the prey's goal is to avoid contact or adjacency with all predators (provided the prey moves). This creates an interesting dichotomy of competition between species (predator and prey), while promoting cooperation, coordination, and collaboration within the predator species. Although competition weighed against cooperation within the predator group has been explored [20], the majority of studies utilize the predator-prey pursuit environment to investigate team dynamics with respect to a shared goal.

The original predator-prey pursuit environment (Figure 1) required four predator agents to surround the single prey agent from four directions in a discretized grid world [1, 9, 21, 23]. The predator agents were guided by an algorithm and their movements were limited to one grid square per time step to an adjacent available square (not occupied by another agent or a boundary) in only the vertical or horizontal directions (no diagonal movements). The prey agent was restricted to the same criteria and guided by random movement. The goal of their simulation experiments was to show the impact that varying degrees of agent cooperation and control had on the efficiency of prey capture. This first rendition of the predator-prey pursuit problem introduced an environment to test the effectiveness of an algorithm to cooperate in a well-constrained domain.

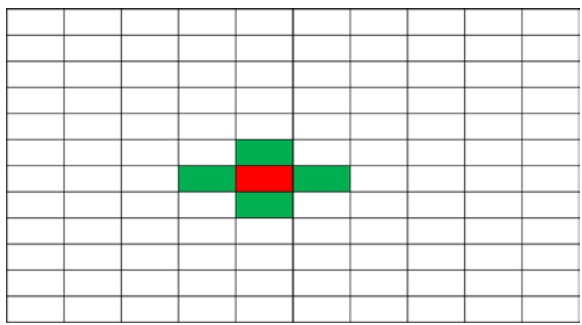

**Figure 1.** Typical predator-prey pursuit environment utilized a discretized grid world bounded on all sides. The goal was to have four predator agents (green blocks) surround the prey agent (red block) on four sides (top, bottom, left, and right). Agents could only move one square at a time and neither could move diagonally nor catch the prey from an adjacent diagonal location.

Cooperative algorithms inherently enable predators to collaborate [12, 24-28]. However, under certain conditions, predators using a greedy strategy may have greater success [29], though cooperative algorithms often win out when a sophisticated prey is faced [23], and in some environments prey benefit most from a mix of greedy and altruistic strategies [20]. These algorithmic explorations of multi-agent cooperation show how the predator-prey pursuit environment provides an ideal testbed for understanding collaborative agent behavior.

Since the origination of the predator-prey pursuit environment, many studies have leveraged manipulations to the pursuit environment to investigate human and artificial intelligence behavior in multiagent systems. The next section explores many of these predator-prey environmental manipulations along with the experimenters' goals to provide a foundation for simulating tactical edge scenarios.

## Modifications to the Predator-Prey Pursuit Environment

Many multiagent research efforts utilize the original discretized predator-prey pursuit environment shown in Figure 1, implementing a toroidal grid world and requiring the single prey to be blocked on all sides by four predators [7, 21, 22]. These studies tend to focus less on the structure of the environment and more on how specific sets of predator strategies impact cooperation and teamwork in homogenous and heterogeneous groups of predators. Whereas this foundational work is important, the remainder of this section will discuss modifications to the predator-prey pursuit environment and their implications.

### Changes in Task Constraints

One of the simplest modifications of the original predator-prey pursuit task is to change the constraints of the task in order to address specific questions or increase task realism. The most straight forward example of this is pursuit environments that are made more complex through the use of continuous spaces. While in grid worlds the prey are typically considered caught when surrounded on four sides by predators (Figure 1), continuous environments require a predator to come within some small distance of [30], or to "tag" (touch or overlap in center of mass) [17, 31] they prey. While discrete environments are typically easier to analyze and reason over, it is worth recognizing that most real-world tasks take place in a continuous domain. Thus, the application of theoretical and empirical revelations that emerge from continuous (or other more realistic) predator-prey pursuit environments will aid our understanding of real life pursuit problems.

As an intermediate step, some authors have extended (while maintaining) the discrete environment with diagonal movements [32], or more complex cell-shapes (such as hexagons [18, 32], or irregular convex cells [33]). The timing of predator and prey movement has also been manipulated, with the standard parametrization allowing agents to move synchronously per time step [32], alternating agents' movements through a sequence of time steps (e.g., turn-taking as described in [34]), or allowing more realistic unrestricted, asynchronous movement [23] such that predators and prey can react to each other at flexible time intervals.

Manipulations to environment bounding have illuminated a dependence between structure and predator pursuit strategies. Unbounded non-toroidal environments have been used, including unbounded planes where pursuers move along an unbounded curvature [35], as well as bounded environments, e.g., [36] where traffic in a police chase is restricted to a closed grid of streets or [24] where encountering the edge of the environment results in death. These manipulations have clear implications for pursuit and evasion strategies, since a non-toroidal unbounded environment will allow fast enough prey to continue indefinitely in a given direction and bounded environments contains corners in which prey can be trapped.

### Changes in the Number of Agents

The most frequent predator-prey pursuit environment modifications have been to vary the number of agents. In one study, the number of pursuers (predators) was varied between one and two in a discrete environment with block obstacles to investigate how different agent learning parameters (Q-learning: learning rate, discount factor, and decay rate), implemented into both the predator and prey (in one condition), alter evader (prey) capture time [19]. In other work, the number of competitive (egoistic) and collaborative (altruistic) predator agents was varied along with the total number of predators (up to 20) to understand how different sizes of homo- and heterogeneous egoistic and altruistic groups of predators catch a single prey [20]. Together, this work demonstrates how simply changing the number of predators and including various types of obstacles can illuminate aspects of collaborative behavior while simultaneously testing the effectiveness of different algorithmic approaches.

### Modifications of the External Environment

Obstacles in the pursuit environment can take on various attributes that force agents to adapt and develop more sophisticated behaviors to achieve the task goal. Typically, these obstacles are static and must be circumnavigated as was described previously [17, 28, 30, 37], but they may take on attributes that disrupt or eliminate an agent (predator or prey) [24]. In one research effort, three pursuers (predators) needed to collaborate in a 2-D environment with complex maze-like obstacles (branching obstacles) to capture an intruder (prey) before it escaped [37]. The goal of this particular study was to identify an optimal strategy that emphasized group reward over individual reward, essentially discovering optimal collaboration within their environmental. Dynamic obstacles have also been utilized to investigate a predator-prey like task where a police chase was modulated with varying degrees of street traffic [36]. In general, these studies found that successful agent behavior was dependent on the attributes of the obstacles.

### Changes in Agent Capabilities

Similar to the inclusion of obstacles in the pursuit environment, restricting how far agents (predators and prey) can see necessitates changes to agent behavior to achieve task success. Some research allows the agents to be omniscient, where all entities' positions are known at every time point [17, 35], whereas other studies permit agents to see in a straight line until an occlusion is encountered [33, 38-40]. Some allow predators and prey to see only within a given range around themselves [21, 30, 41] or apply random limitations to predator vision [42]. While some research explores predators with limited sensing abilities, these studies allow information to be shared among predators [21, 36], either through direct communication (e.g., police radio, [36]) or indirectly (e.g., by leaving cues such as pheromone trails in the environment) [26]. Other work has explored an imbalance between sensing abilities of the predators and prey. For example, a predator may see a prey from a greater distance than the prey can detect, reflecting a more realistic scenario where the predator is on the hunt for a non-vigilant prey [21]. Other work has further modified agent vision or sensing abilities by testing something akin to sound, where a predator agent can hide around corners [43]. Together, these studies show the importance of agents' sensing capabilities and how manipulations illuminate the dependence between pursuit strategies and these capabilities.

Limitations to agents' sensing capabilities in simulation experiments naturally extends the pursuit problem into the physical domain with robots. A common physical limitation of a robotic agent is the field of view [44]. The field of view is dependent on two main factors, 1) the distance between the visual system and the object, and 2) the degrees of visual angle the system can see (the average human can see approximately 210 degrees of visual angle). Given that robotic systems need to have some visual representation of the environment around them for obstacle avoidance and navigation, certain limitations to field of view can cause a catastrophic failure (possibly disabling or destroying the robot). Therefore, it is of critical importance to understand how manipulations to field of view effect robotic systems' pursuit behavior.

### Additional Dimensions

In general, 3-D environments provide an opportunity for more complex behavioral strategies, such as those required in aviation, aquatics, and for traversing uneven surfaces [31, 39]. Through investigation of these various 3-D environments, a natural emergence of behavioral strategies that depend on the terrain can be discovered. A physical example of this phenomenon is found in the attack and evasion strategies for spiders and crickets as forest leaf litter change geometries between winter and summer [45].

## Adaptation of the Predator-Prey Pursuit Environment to Tactical Edge Scenarios

The predator-prey pursuit environment has been used by researchers since 1985 [1] to investigate a host of topics including various aspects of group dynamics [25, 46], pursuit strategies [3, 47], escape/evasion strategies [27, 41, 48], and multi-agent systems [18-20, 49] to name a few. The last portion of this article discusses a set of proposed manipulations to the predator-prey pursuit environment to investigate potential simulated tactical scenarios that easily map to a physical domain.

### Impact of Spatial Constraints on Strategies and Behavior

The predator-prey environment size, or the space available for agents (predators and prey) to move around in, has been manipulated to understand how environmental geometries [18, 32] and sizes [50] influence agent behavior in discrete spaces. Building upon this body of work and extending into the continuous domain [17], it would be of interest to start with a trivial minimized space, such that the predators always catch the prey within a short period of time, and incrementally increase the size of the environment to gain a quantitative understanding of how environment size impacts prey catch times as a function of increasing environment size. We would expect that prey catch time would increase with a power function of the dimensions of the environment, with a certain size resulting in a near-zero probability of catch. The dimensional expansion can provide a fundamental understanding of task difficulty with respect to environment size. In a simulated tactical scenario, such an understanding would allow us to better assess the probability of acquiring a moving target given the estimated escape space available. This could also lead towards valuable insight into the type of agent capabilities needed to estimate a high probability of mission success (e.g., small and fast reconnaissance agents).

## Modifying Predator Capabilities

While holding all other possible manipulations constant, modifications to agent velocity or rate of predator agent movement in a continuous space introduces many additional degrees of freedom and provides an opportunity to simulate tactical teams with homo- and heterogeneous capabilities. A systematic sweep through a range of velocities for a set of predator agents can provide a valuable mapping between agent velocity and mission success (i.e., prey catch time). However, it is important to note that changes to predator velocity are relative to prey velocity and should likely be thought of as a ratio. With that stated, a simultaneous change to all predator agents' velocities (increase or decrease) relative to the prey agent will result in an understanding of the relationship between homogeneous alterations to a team's capabilities and mission success. Similarly, heterogeneous manipulations to the predator agents' velocities relative to the prey agent in a continuous predator-prey pursuit environment can provide an estimate of mission success for the various capabilities (velocity sweep across all predators) in this simulated target acquisition tactical scenario.

On the other side of the proverbial coin, holding predator attributes constant and applying manipulations to the prey's velocity introduces additional dimensionality to the task domain and may result in the need for more complex collaborative behavior to achieve predator agent mission success. As was suggested for the predator agents, a velocity sweep for the prey would dictate task difficulty (easy for low velocities and hard for high velocities) and should result in a spectrum of competitive to collaborative predator agent behaviors for low to high velocities respectively. It is important to note that modifications to the prey's velocity would need to be relative to the predator agents' velocities. We would expect that manipulations from low to high predator to prey velocity ratio would result in the shifts from easy to hard for task difficulty and competitive to collaborative predator agent team behavior. These simulation experiments might represent the differences between having a homo- or heterogeneous teams of slow, heavy, powerful assets (e.g., tanks), camouflaged insurgent ground assets, or drones in reconnaissance or target acquisition scenarios.

## Modifying Prey Capabilities

In the prey manipulation domain, introducing multiple prey in various forms could change the task goals entirely. The inclusion of a second prey expands the dimensionality of the task domain to include homo- and heterogeneous adversarial dynamics (two prey agents forming an adversarial team), a potential decoy (catching the decoy prey agent does not complete the mission), and variable temporal mission objective windows (the predator agents must coordinate to catch both prey agents within a preselected duration). The inclusion of additional prey agents (number of prey agents > 2) can further increase the complexity of the task domain, possibly to the point of which the probability of simulated tactical mission success goes to zero. Mission failure is important to explore, especially in simulated environments, in order to maximize the probability of mission success in the multi-domain battlespace.

Modifications to coordinate locations in the simulated environment, while holding all agent attributes (both predator and prey) constant, permits an investigation of degraded agent capabilities. Coordinate manipulations can take the form of static impassible barriers that represent buildings/walls/obstacles, patches that induce injury by temporarily (short duration) or permanently (remaining duration) reducing/minimizing/stopping agents' (predator and/or prey) movements or small environmental regions with simulated hidden explosives that completely remove an agent from further participation in the mission. Other coordinate manipulations are possible (e.g., teleport agents randomly around the environment), but they might not have an easily identifiable correspondence to tactical scenarios. Therefore, environmental manipulations that easily map to tactical scenarios can provide an estimate of the relationship between agent capability degradation and mission success.

## Conclusion

Complexification of the predator-prey pursuit environment to include modifications that easily map to simulated tactical scenarios allows for the adaptation of computational agents to these domains. This line of predator-prey pursuit research can be extended to a physical environment that accommodates the testing of robotic platforms working with Soldiers in target acquisition training drills, for the eventual implementation on the multi-domain battlefield.

## Acknowledgements

Research was sponsored by the Army Research Laboratory and was accomplished under Cooperative Agreement Number W911NF-17-2-0003. The views and conclusions contained in this document are those of the authors and should not be interpreted as representing the official policies, either expressed or implied, of the Army Research Laboratory or the U.S. Government. The U.S. Government is authorized to reproduce and distribute reprints for Government purposes notwithstanding any copyright notation herein.